\documentclass[a4paper]{article}

\usepackage{INTERSPEECH2022}
\usepackage{flushend}
\usepackage{multirow}
\usepackage[skip=8pt]{caption}
\usepackage{subcaption}
\usepackage{makecell}

\title{Efficient Transformer-based Speech Enhancement Using Long Frames and STFT Magnitudes}

\name{Danilo de Oliveira, Tal Peer, Timo Gerkmann}
\address{Signal Processing (SP), Universit\"at Hamburg, Germany}
\email{\{danilo.oliveira, tal.peer, timo.gerkmann\}@uni-hamburg.de}

\begin{document}

\maketitle
\begin{abstract}
The SepFormer architecture shows very good results in speech separation. Like other learned-encoder models, it uses short frames, as they have been shown to obtain better performance in these cases. This results in a large number of frames at the input, which is problematic; since the SepFormer is transformer-based, its computational complexity drastically increases with longer sequences.
In this paper, we employ the SepFormer in a speech enhancement task and show that by replacing the learned-encoder features with a magnitude short-time Fourier transform (STFT) representation, we can use long frames without compromising perceptual enhancement performance. We obtained equivalent quality and intelligibility evaluation scores while reducing the number of operations by a factor of approximately 8 for a 10-second utterance.
\end{abstract}
\noindent\textbf{Index Terms}: speech enhancement, attention, transformer
\section{Introduction} \label{introduction}
The tasks of speech enhancement and general audio source separation have greatly benefited from advances in neural networks in the past decade.  Methods for speech enhancement have traditionally used statistical characteristics of noise and speech to estimate the spectrum of the noise and subsequently of clean speech \cite{gerkmannvincent2018, noisereducitonsurvey}. However, such methods usually suffer in non-stationary noise conditions, which can be remedied using machine learning methods. Architectures with recurrent neural networks like LSTMs \cite{hochreiter_long_1997}, the attention mechanism \cite{bahdanau_neural_2015} and temporal convolutional networks \cite{van_den_oord_wavenet_2016} have showcased interesting performance improvements in recent years \cite{vincent_speech_2015, tan_convolutional_2018, kim_t-gsa_2020, richter2020, rehr2021}.

A neural network-based approach that has shown good results and flourished with the increased capacity and computational power of modern processors is learned-domain speech processing, that is, training of models that learn representations of the audio inputs and perform processing steps on them. This is done in an end-to-end manner and without hand-crafted, fixed transforms like the short-time Fourier transform (STFT). These methods were popularized with the publication of Conv-TasNet \cite{luo_conv-tasnet_2019}, which fostered research and preceded many other works on end-to-end encoder/masker/decoder architectures.

Nevertheless, later contributions have shown that the good performance obtained by Conv-TasNet does not come specifically from the freely-learned convolutional encoder/decoder pair. In \cite{ditter_multi-phase_2020}, equivalent results were obtained by replacing the learned encoder with a multi-phase gammatone analysis filterbank. In \cite{heitkaemper_demystifying_2020}, the authors show that gains from Conv-TasNet can be attributed to the high time resolution and the time-domain loss.

While learned-domain methods have excellent performance and the benefit of not decoupling magnitude from phase, they usually work on short frames (2ms), which implies having to deal with a larger number of frames if compared to traditional STFT frame sizes ($\sim$32ms). This is especially problematic in models using the attention mechanism, whose computational complexity grows quadratically with respect to the sequence length due to the dot-product operation.

Dual-path methods have managed to alleviate some issues related to the modeling of long sequences for speech applications: in \cite{luo_dual-path_2020} the authors proposed segmenting the sequence of frames into chunks and processing the sequences with LSTMs inside the chunks, followed by processing across chunks, being therefore able to model both short- and long-term dependencies with a reduced computational footprint; the authors of \cite{chen_dual-path_2020} proposed using what they called the \textit{improved Transformer}, using multi-head self-attention along with LSTMs; and in \cite{subakan_attention_2021} the SepFormer model was introduced, relying on attention only, as in the original Transformer paper \cite{vaswani_attention_2017}.

Although dual-path methods reduce complexity to a feasible level, the number of frames is still cumbersome, requiring large amounts of memory during training. Another drawback common to existing learned-domain approaches is that the models usually work with 8kHz audio data, which is a considerable disadvantage against wideband methods. Additionally, the learned-encoder features have reduced interpretability compared to well-established, fixed filters such as the STFT. Therefore, using larger frame time-frequency representations still presents itself as a desirable feature, though working with complex representations brings additional challenges.

Studies on the contribution of STFT magnitude and phase spectra in speech processing \cite{kazama_significance_2010, pariente_filterbank_2020,peer_intelligibility_2021,peer_phase-aware_2022} have shown that the relative importance of phase varies considerably with frame size. In particular, the loss of spectral resolution renders the magnitude less relevant at very short frames ($\leq$ 2ms). This does not apply to the phase spectrum, which encodes information related to the zero-crossings of the signal \cite{kazama_significance_2010}. For larger frames (around 32ms), the opposite holds: magnitude is more important than phase \cite{wang_unimportance_1982}. These effects are also reflected in enhancement/separation evaluation metrics. Based on these observations, with an appropriate choice of frame size we can easily port existing attention-based models to use time-frequency representations by only processing the magnitudes. In this paper we investigate what compromises and benefits can be attained when working with magnitudes of longer frames.

\section{Speech enhancement} \label{speech_enhancement}

The task of speech enhancement is formulated in this work as follows: we consider a mixture $x$ consisting of clean speech $s$ and additive noise $v$ at time samples $n$:

\begin{equation}
    x(n) = s(n) + v(n).
    \label{mixture}
\end{equation}

Except where explicitly needed, we will drop the indexes, and represent the signals as vectors (in bold). The estimated clean signal $\mathbf{\hat{s}}$ is obtained via masking of the noisy input. First the time-domain signal is encoded into a representation more adequate for separation:

\begin{equation}
    \mathbf{w} = \textrm{Encoder}(\mathbf{x}) \,,
    \label{encoding}
\end{equation}
then a mask $\mathbf{m}$ is applied to this encoded representation
\begin{equation}
    \mathbf{d} = \mathbf{m} \odot \mathbf{w} \,,
    \label{masking}
\end{equation}
where $\odot$ denotes the element-wise multiplication operator. We then apply a decoder to the masked representation in order to return to the time domain, resulting in the clean speech estimate
\begin{equation}
    \mathbf{\hat{s}} = \mathrm{Decoder}(\mathbf{d}) \,.
    \label{decoding}
\end{equation}
This process is illustrated in Figure~\ref{fig:encdec}.
\section{Architecture} \label{architecture}
\begin{figure}
    \begin{subfigure}[b]{0.5\textwidth}
        \centering
        \includegraphics{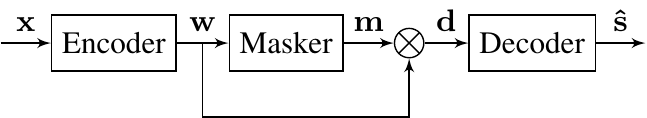}
        \caption{General architecture.}
        \label{fig:encdec_base}
    \end{subfigure}
    \par\medskip
    \begin{subfigure}[b]{0.5\textwidth}
        \centering
        \includegraphics{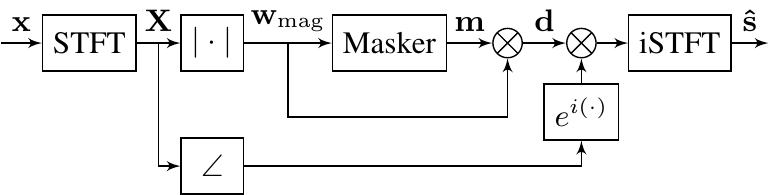}
        \caption{Magnitude spectrogram architecture.}
        \label{fig:encdec_mag}
    \end{subfigure}
    \caption{Encoder/Masker/Decoder structures.}
    \label{fig:encdec}
\end{figure}

The models in this work are organized in the structure of an encoder, a masking part and a decoder, as depicted in Figure~\ref{fig:encdec_base}. Experiments were conducted using either a freely-learned encoder/decoder pair, or the STFT as encoder and the inverse STFT (iSTFT) as the decoder.

\subsection{STFT encoder/decoder}
As presented in Section~\ref{introduction}, a commonly used representation for audio signals is the time-frequency domain, which captures more prominently the structure of the audio signals and also facilitates source separation tasks due to increased sparsity. The method used in this work for obtaining the time-frequency representation $X\in\mathbb{C}^{F\times L}$ is the STFT, obtained by applying the discrete Fourier transform (DFT) to windowed, overlapped frames:
\begin{equation}
    X(f,l) = \sum_{n=0}^{M-1} x(n + lH) w(n)e^{-j\frac{2\pi kn}{M}} \,,%, \quad k \in \{0, ..., K-1\}
    \label{stft}
\end{equation}
where $f$ and $l$ are the frequency bin and frame indexes, respectively, $n$ is the local time index, $w(n)$ is a window function, $M$ is the window length and $H$ is the hop length. We use the one-sided DFT, so with this type of encoder we have $F = \frac{M}{2} + 1$. We can define the frame overlap ratio as $\frac{M-H}{M}$, which in this work will be given as a percentage.
The input to the masker in this case is the magnitude of the complex time-frequency representation:
\begin{equation}
    \mathbf{w_\mathrm{mag}} = |\mathbf{X}| \,.
    \label{stft_enc}
\end{equation}
The masked magnitudes processed by the masker are joined with the with the noisy phase $\arg(\mathbf{X})$ and then fed to the decoder, which is the inverse STFT operation. This is shown in Figure~\ref{fig:encdec_mag}.
\subsection{Learned encoder/decoder}
In the case of a learned encoder/decoder pair, similar to \cite{luo_conv-tasnet_2019, subakan_attention_2021}, we use one-dimensional convolutional layers (Conv1d). The encoder transforms the audio signals directly into a higher dimensional representation, and a rectified linear unit (ReLU) enforces a non-negativity constraint, so we have
\begin{equation}
    \mathbf{w_\mathrm{conv}} = \mathrm{ReLU}(\mathrm{Conv1d}(\mathbf{x})) \,,
    \label{conv_enc}
\end{equation}
where $\mathbf{w_\mathrm{conv}} \in \mathbb{R}^{F\times L}$. The decoder uses a transposed one-dimensional convolutional layer (ConvTranspose1d) to convert the masked estimate $\mathbf{d}$ back to the time domain:
\begin{equation}
    \mathbf{\hat{s}} = \mathrm{ConvTranspose1d}(\mathbf{d}) \,.
    \label{conv_dec}
\end{equation}
\subsection{Masker network}
\begin{figure*}
    \begin{subfigure}[b]{\textwidth}
        \centering
        \includegraphics[scale=0.9]{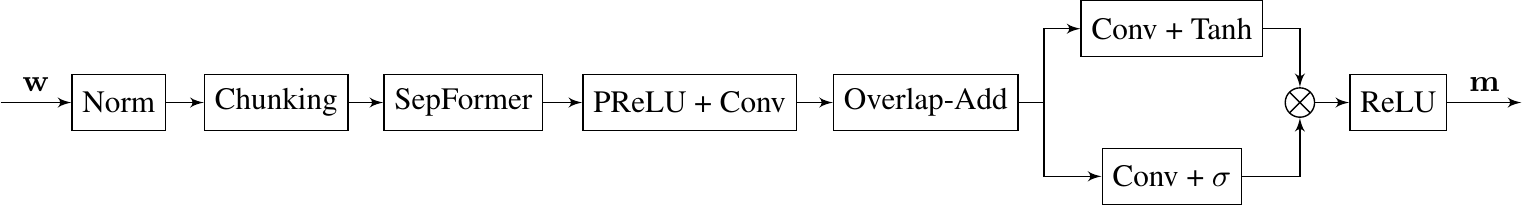}
        \caption{Overlap-add chunking and mask generation.}
        \label{fig:dual_path}
    \end{subfigure}
    \par\medskip
    \begin{subfigure}[b]{\textwidth}
        \centering
        \includegraphics[scale=0.9]{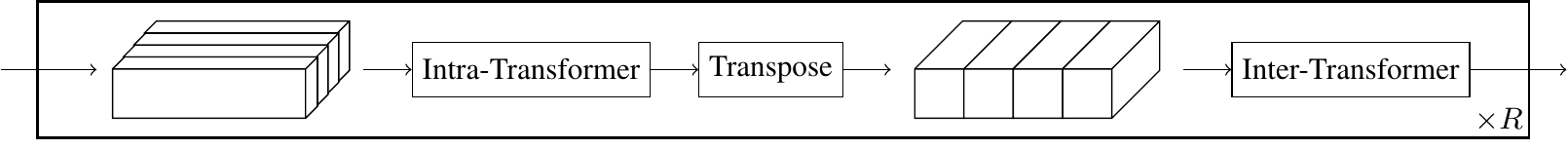}
        \caption{SepFormer architecture, with the dual-path scheme.}
        \label{fig:sepformer}
    \end{subfigure}
    \par\medskip
    \begin{subfigure}[b]{\textwidth}
        \centering
        \includegraphics[scale=0.9]{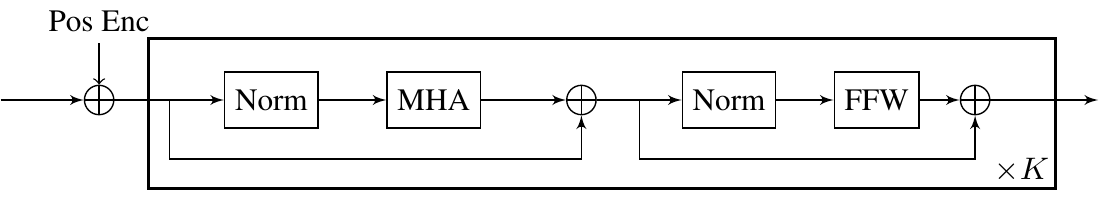}
        \caption{Transformer block structure used for both the Intra- and the Inter-Transformer.}
        \label{fig:transformer}
    \end{subfigure}
    \caption{Architecture of the system}
    \label{fig:system}
\end{figure*}

The masker DNN in our experiments is a reduced version of the SepFormer \cite{subakan_attention_2021}, based on Huang's implementation \cite{huang_stabilizing_2021}. This architecture is displayed in Figure~\ref{fig:system} and follows the dual-path principle introduced in \cite{luo_dual-path_2020}: the mixture frames are chunked into overlapping segments, which are then stacked. As can be seen in Figure~\ref{fig:sepformer}, the sequence modeling steps are applied first along the chunks (intra-chunk processing) and after that the dimensions are transposed and the sequence processing is executed across chunks (inter-chunk). This process is repeated $R$ times. 

Following \cite{subakan_attention_2021}, the intra- and inter-chunk processing is done without recurrent neural networks, using instead only a sequence of $K$ transformer blocks. These transformer blocks include a multi-head attention (MHA) stage and a feed-forward (FFW) block, both preceded by layer normalization (Norm) steps and with skip connections. Positional encoding is added at the blocks' inputs to introduce position information to the model. This is displayed in Figure~\ref{fig:transformer}. The output of the SepFormer block is further processed by parametric ReLU (PReLU) and Convolutional (Conv) layers, and the chunks are merged back via the overlap-add method.

Figure~\ref{fig:dual_path} illustrates the masking principle used, the same as presented in \cite{chen_dual-path_2020}, with sigmoid ($\sigma$) and hyperbolic tangent (Tanh) branches performing gating and limiting the masks to the interval $[-1,1]$. A ReLU layer follows this step, enforcing coherence with the non-negative inputs to be masked.

The attention mechanism has quadratic complexity with respect to the sequence length $L$, which can be a problem when using short frames: with the number of frames being given by $L = 1 + \lfloor(N-M)/H\rfloor$, where $N$ is the total number of samples, a system with short window size $M$ will result in a large number of elements to process. The dual-path architecture is able to reduce the complexity from $\mathcal{O}(L^2)$ to $\mathcal{O}(L\sqrt{L})$ in the best case scenario, when the chunk size is $C = \sqrt{L}$. However, this only mitigates the problem to a certain extent; as the number of frames starts becoming much larger than the chunk size, the inter-chunk processing stage starts to dominate, with the complexity tending to quadratic again.
\section{Experiments} \label{experiments}
\subsection{Dataset}
The models were trained on the DNS-Challenge dataset \cite{reddy_interspeech_2020}. We generated 100 hours of 4 second long noisy mixtures sampled at 16kHz, with 20\% reserved for validation. Testing was performed on clean samples from a subset of the WSJ0 \cite{garofolo_csr-i_1993} corpus mixed with noise from the CHiME3 Challenge dataset \cite{barker_third_2017}, at SNRs ranging from -10dB to 15dB, at 5dB intervals.
\subsection{Model}

The number of blocks used in the SepFormer was $R=2$, $K_\text{intra}=4$ and $K_\text{inter}=4$ for the number of repetitions of the attention mechanism in the intra- and inter-chunk transformer blocks, respectively. The number of dimensions used in the feed forward layers was 256. The chunking was performed with 50\% overlap.

In the case of the convolutional encoder/decoder pair, the number of learned filters was 256. The filter size was set to 32 (or 2ms at 16kHz), with stride 16, therefore 50\% overlap. For the STFT case, the window function is the Hann window, with length 512 (or 32ms at 16kHz), with 50\% or 75\% overlap. All configurations of the model contain approximately 6.6 million parameters.

For model training, ADAM \cite{kingma_adam_2015} was used as the optimizer, with a learning rate of $10^{-3}$, halved after 5 epochs without improvement. Gradient clipping at 5 was employed. We did not use dynamic mixing during training. Following the original SepFormer paper, the loss function used for training was the scale-invariant signal-to-distortion ratio \cite{roux_sdr_2019}.
\section{Results and discussion} \label{results}
\subsection{Enhancement performance}

The estimated utterances were evaluated on instrumental perceptual metrics: POLQA \cite{beerends_perceptual_2013} for speech quality and ESTOI \cite{jensen_algorithm_2016} for intelligibility. The results of the different configurations are organized in Table~\ref{tab:results}.

In the learned-domain case, the chunk size of 250 as in \cite{subakan_attention_2021} performs best against a setup with shorter chunks, hinting at the importance of modeling short-term relations in the sequence. Nevertheless, long-term relations also play a role in performance, as can be seen in the magnitude STFT experiments: the configuration with chunks size 50 seems to find a balance between short- and long-term, if compared to the models with 25 and 100. Setting the frame overlap to 50\% to obtain even fewer frames resulted in degradation to the quality metrics. The performance of the STFT model at different input SNRs remained consistent with the learned-encoder case. Informal listening evaluations confirmed the findings from the instrumental scores and found the learned-encoder estimates to contain a buzzing sound that is absent from the magnitude STFT outputs. Audio examples are available online\footnote{\texttt{https://uhh.de/inf-sp-magnitudetransformer}}.
\subsection{Execution profiling}
\begin{table*}[th]
  \caption{Model configurations and their respective computational performance and speech enhancement scores. GMAC and execution time profiling was computed for an input containing 10 seconds of audio. Processing times were averaged over 10 executions.}
  \label{tab:results}
  \centering
  \tabcolsep=0.15cm
  \scalebox{0.95}{
  \begin{tabular}{c|c c c|c c c|c c}
    \textbf{Model} & 
        \textbf{\makecell{Frame\\(ms)}} &
        \textbf{\makecell{Frame\\Overlap}} &
        \textbf{\makecell{Chunk\\Size}} &
        \textbf{GMACs} &
        \textbf{\makecell{GPU\\Time (ms)}} &
        \textbf{\makecell{CPU\\Time (ms)}} &
        \textbf{POLQA} &
        \textbf{ESTOI} \\
        
    \hline\hline
    \multirow{2}{*}{Learned-domain SepFormer \cite{subakan_attention_2021}} & 2  & 50\% & 250 & 45.75 & 69 & 909 & 2.98          & \textbf{0.79} \\
                                                                            & 2  & 50\% & 100 & 45.10 & 66 & 895 & 2.91          & \textbf{0.79} \\
    \hline
    \multirow{4}{*}{Magnitude STFT SepFormer (ours)}                        & 32 & 75\% & 100 & 6.26  & 11 & 109 & 2.92          & 0.77          \\
                                                                            & 32 & 75\% & 50  & 5.93  & 11 & 153 & \textbf{3.01} & 0.78          \\
                                                                            & 32 & 75\% & 25  & 5.99  & 11 & 106 & 2.95          & 0.78          \\
                                                                            & 32 & 50\% & 25  & 3.08  & 11 & 89  & 2.78          & 0.76          \\
    \hline
  \end{tabular}
  }
\end{table*}

The metrics concerning execution are also given in Table~\ref{tab:results}, in the form of the number of giga multiply–accumulate operations (GMACs) and the execution time for a 10-second input, averaged over 10 executions. The profiling was executed on a computer equipped with an Intel Core i9-10900X CPU at 3.70GHz and an NVIDIA GeForce RTX 2080 Ti graphics card. 

In terms of performance profiling, the chunk size does not play a big role in offline processing, as larger chunks will increase computation efforts for the intra-chunk step but reduce it for the inter-chunk part, and the opposite happens with shorter chunks. Frame size and overlap, however, have a more important impact on execution, since they control the number of elements to process. Going from a learned-encoder setup with 2ms frames overlapped at 50\% to an STFT configuration of 32ms frames with 75\% overlap, we can see a significant reduction of complexity. Taking the best performing model from each category (chunk size of 250 for the learned encoder case and 50 for its STFT counterpart), we can see a reduction of the number of accumulate-add operations by a factor of 7.7. This reduction also translates to execution time, which is reduced approximately by a factor of 6, for both GPU and CPU inference.

\begin{figure}[ht]
\includegraphics{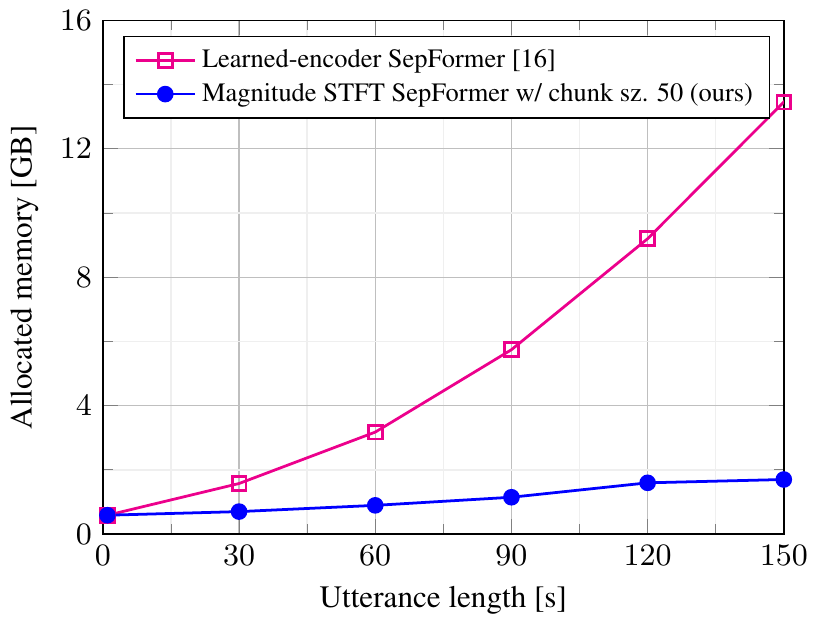}
\caption{Maximum memory allocation of the script performing the forward pass for increasing utterance lengths, running on CPU. The learned-encoder SepFormer uses chunk size 250, frame size 2ms and 50\% overlap, whereas the magnitude STFT SepFormer uses 50, 32ms and 75\%.}
\label{fig:plot_mem}
\end{figure}

Looking at further practical aspects of a speech enhancement model implementation, memory allocation was profiled on CPU, with the results presented in Figure~\ref{fig:plot_mem}. The learned-encoder variant allocates around 2GB of RAM shortly after 30 seconds of input, and over 4GB after the 60 second mark. For low-resource, embedded devices, usage of such a model is therefore severely hindered. For the whole range of values analyzed, up to 2.5 minutes of uninterrupted data, the magnitude STFT SepFormer kept memory usage below 2GB.

Considering a potential application in an online speech enhancement scenario, chunk processing times have to be analyzed. The time it takes for a new chunk to be complete for processing acts as a maximum value for feasible online execution. Taking into account the fact that intra-chunk attention calculations have complexity linked to the fixed chunk size and assuming the attention matrices for previous chunks can be stored to save operations, the most important step to be examined is the inter-chunk step. 

\begin{figure}[ht]
\includegraphics{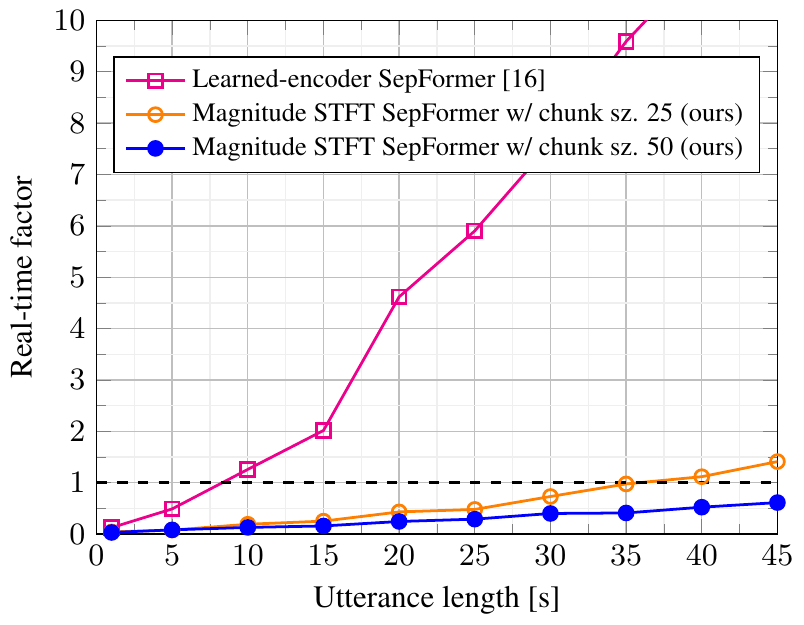}
\caption{Real-time factor (RTF) for increasing utterance lengths. RTF is defined here as the chunk processing time divided by the time it takes for a chunk to be complete and ready to be processed (it is different for different setups). The dashed line indicates $RTF=1$, the point where the required processing time starts getting longer than the chunk length plus the shift. The learned encoder works with 2ms at 50\% overlap, with chunk size 250. For the magnitude STFT, these values are 32ms, 75\% and 25 or 50, respectively.}
\label{fig:plot_time}
\end{figure}

For practical online operation, the time it takes to process a chunk must be shorter than its length plus its shift. 
Figure~\ref{fig:plot_time} shows that in the learned-encoder case this threshold is already reached for sequences shorter than 10 seconds, whereas the magnitude STFT version in the configurations tested kept the execution below the threshold for at least 35 seconds. Note that due to frame length and overlap configuration, the algorithmic latency of the STFT encoder in the best performing configuration (chunk size of 50) is double the value for the learned encoder. This makes it unsuitable for real-time communication applications, but can be remedied by reducing the chunk size or increasing its overlap, at the expense of slight degradation in speech enhancement and computation time. We therefore included in Figure~\ref{fig:plot_time} the magnitude model using a chunk size equal to 25, with an algorithmic latency similar to the learned-encoder case. It is also worth mentioning that other blocks in the model add processing time, such as the encoder and the chunking step. Their contribution to the overall complexity is nevertheless not as significant as the inter-chunk attention.

\section{Conclusion} \vspace{-2px}\label{conclusion}
Recent advances in learned-domain speech processing show excellent performance, but the models are demanding in terms of memory and processing power, therefore lacking feasibility in practical applications. Motivated by previous contributions on learned and traditional filterbanks and on the relation between frame size and magnitude/phase processing, we show that by replacing the learned features with STFT magnitudes, we can obtain equivalent performance in terms of perceptually-motivated metrics while considerably reducing resource allocation and processing time. These findings are a big step towards making the implementation of state-of-the-art transformer-based speech enhancement systems possible in real-life applications, especially on embedded devices.
\section{Acknowledgements}\vspace{-2px}
This work has been funded by the Deutsche Forschungsgemeinschaft (DFG, German
Research Foundation) -- project number \texttt{247465126}. We would like to thank J. Berger and Rohde\&Schwarz SwissQual AG for their support with POLQA.
\bibliographystyle{IEEEtran}
\clearpage
\atColsBreak{\vskip5pt}
\bibliography{references,mybib}

\end{document}